\documentclass{emulateapj}
\usepackage{times}

\def\la{\mathrel{\mathpalette\fun <}}
\def\ga{\mathrel{\mathpalette\fun >}}

\def\simpropto{\lower.2ex\hbox{$\; \buildrel \sim \over \propto \;$}}

\def\fun#1#2{\lower0.837ex\vbox{\baselineskip0ex\lineskip0.209ex
  \ialign{$\mathsurround=0ex#1\hfil##\hfil$\crcr#2\crcr\sim\crcr}}}

\def\msun{M_\odot}
\def\msunyr{M_\odot \ {\rm yr}^{-1}}

\def\sles{\lower2pt\hbox{$\buildrel {\scriptstyle <}
   \over {\scriptstyle\sim}$}}

\def\sgreat{\lower2pt\hbox{$\buildrel {\scriptstyle >}
   \over {\scriptstyle\sim}$}}

\def\la{\mathrel{\mathpalette\fun <}}
\def\ga{\mathrel{\mathpalette\fun >}}

\def\msun{M_\odot}
\def\msunyr{M_\odot \ {\rm yr}^{-1}}

\begin{document}

\title{The {\it Kepler} light curves of V1504 Cygni 
            and V344 Lyrae: A study of
                       the Outburst Properties}
\shortauthors{CANNIZZO ET AL.}
\author{
 John~K.~Cannizzo\altaffilmark{1,2},
 Alan~P.~Smale\altaffilmark{3},
 Matt~A.~Wood\altaffilmark{4},
 Martin~D.~Still\altaffilmark{5,6},
 Steve~B.~Howell\altaffilmark{ 5}
   }
\altaffiltext{1}{CRESST and Astroparticle Physics Laboratory
               NASA/GSFC, Greenbelt, MD 20771, USA;
                 e-mail: 
               John.K.Cannizzo@nasa.gov}
\altaffiltext{2}{Department of Physics, University of Maryland,
              Baltimore County, 1000 Hilltop Circle,
              Baltimore, MD 21250, USA}
\altaffiltext{3}{NASA/Goddard Space Flight Center,
               NASA/GSFC, Greenbelt, MD 20771, USA}
\altaffiltext{4}{Department of Physics and Space Sciences,
                Florida Institute of Technology,
                 150 W. University Blvd., Melbourne, FL
                  32901, USA}
\altaffiltext{5}{NASA Ames Research Center, Moffett Field, 
                 CA 94035, USA}
\altaffiltext{6}{Bay Area Environmental Research Institute, Inc.,
                560 Third St. West, Sonoma,
                CA 95476, USA}

\begin{abstract}
We examine the {\it Kepler} light curves of V1504 Cyg and V344 Lyr,
                   encompassing $\sim$736 d  at 1 min cadence.
  During this span
          each system exhibited $\sim$$64-65$ outbursts,
   including six superoutbursts.
     We find that, in both systems,
   the normal outbursts lying
  between two superoutbursts increase in duration 
     over time
 by a factor $\sim$1.2$-$1.9,
  and then reset to a small value after the following  superoutburst.
 In both systems
   the trend of quiescent intervals
between normal outbursts is to increase  to a local
maximum about half way through the supercycle
    $-$
   the interval from one superoutburst to the next
    $-$
   and then
to decrease back to a small value by the time of the
  next superoutburst.  
  This is inconsistent with 
Osaki's thermal-tidal model, which predicts a
   monotonic
   increase in the quiescent intervals between
normal outbursts during a supercycle.
  Also, most of the  normal outbursts have an asymmetric,
fast-rise/slower-decline shape, which would be consistent
with outbursts triggered at large radii.
   The exponential rate of decay of the 
   plateau phase of the superoutbursts
  is 8 d mag$^{-1}$ for V1504 Cyg and
    12 d mag$^{-1}$ for V344 Lyr.
 This time scale gives a direct measure
 of the viscous time scale in the outer accretion disk
given the expectation that the entire disk is in the hot,
viscous state during superoutburst.
  The resulting constraint on
the Shakura-Sunyaev parameter, $\alpha_{\rm hot}\simeq0.1$,
  is consistent with the value inferred from the
  fast dwarf nova decays.
  By looking at the slow decay rate for superoutbursts,
  which occur in systems
below the period gap, in combination with 
   the slow decay rate  in one long outburst
above the period gap (in U Gem), we infer
  a steep dependence of the 
 decay rate on orbital period for long outbursts.
     We argue that this relation implies a steep dependence
of $\alpha_{\rm cold}$ on orbital period, which may be consistent
with recent findings of Patterson, and is consistent with tidal
   torquing as being the dominant angular  momentum transport
  mechanism in quiescent disks in interacting binary systems.
\end{abstract}

%

\section{Introduction}

Cataclysmic variables (CVs) are
  semi-detached interacting binaries in which a
  Roche lobe filling K or M secondary transfers 
          matter to a white dwarf (WD).
  CVs evolve to shorter orbital periods
   and show a ``gap''
between $P_h=2$ and 3 (where $P_h = P_{\rm orbital}/1$ hr)
    during which
  time the 
  secondary star loses contact with its Roche lobe
  and mass transfer ceases.
  Thus the binary becomes fully detached.
  At $P_h=2$ the secondary  comes back into 
contact with its Roche lobe and mass transfer resumes.
   For $P_h < 2$ angular momentum loss
  from the binary is
   thought to be due solely to gravitational radiation.
The CV subclass of dwarf novae (DNe) are characterized 
   by their semi-periodic 
   outbursts. 
  SU UMa stars are DNe lying below the period gap
  that 
 exhibit  short, normal outbursts (NOs) and superoutbursts (SOs).
 We refer the time from one SO to the next as the supercycle.
    SOs show superhumps which are modulations in the light curve
at  periods slightly exceeding the orbital  period.
 There are two further subdivisions within the SU UMa grouping:
 (i)  the VW Hyi stars at long orbital periods,
   near $P_h=2$,  for which the
   decay rate is fairly constant during a SO,
  and 
(ii) the WZ Sge stars at short orbital periods,
    a little greater than $P_h=1$,
    which have less frequent, larger amplitude SOs,
   for which the decay rate decreases during a SO.
   DNe outbursts are thought to be due to a limit cycle accretion disk
  instability (Lasota 2001)
   in which material is accumulated in quiescence
  and then dumped onto the WD during outburst.
 During short outbursts
   in longer period DNe,
  a  few percent of the stored mass is accreted, and during long outbursts
   a significant fraction $\sim$0.2 of the stored mass is accreted.
         For the SU UMa stars, a SO  is thought to accrete   $\ga0.7-0.8$
  of the stored mass.  
  Although the accretion disk is never in steady state 
   during the limit cycle, it is close 
   to steady state during SO,
 with the local rate of accretion inside the disk $|{\dot M}(r)|$
   decreasing linearly from a maximum value at the disk inner edge
  to zero at the outer edge. The accretion
  disk modeling has traditionally been done
   within the 
   Shakura \& Sunyaev (1973,
       hereafter SS) formalism, using two values
  for the $\alpha$ viscosity parameter, $\alpha_{\rm hot}$ 
  for gas in the hot, ionized disk, and $\alpha_{\rm cold}$
   for gas in the quiescent disk.

\begin{figure*}
\centering
\epsscale{1.0}
\includegraphics[scale=0.75]{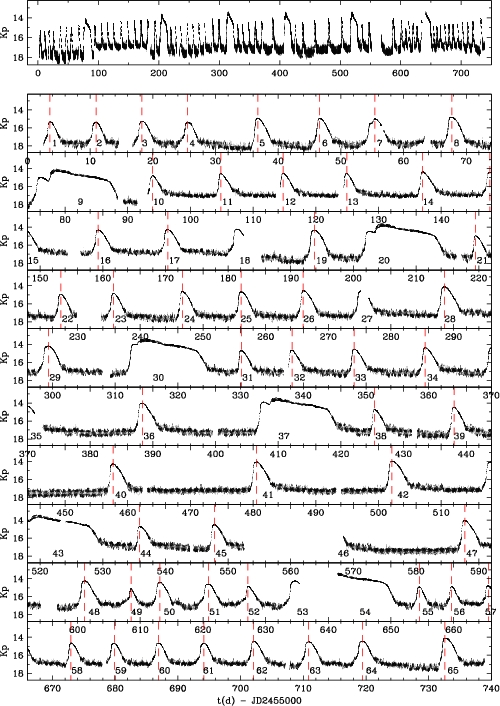}
\vskip .35cm
\figcaption{
  {\it Kepler} light curve for V1504 Cyg 
   at 1 min cadence, spanning $\sim$736 d.
   The light curve covers six  superoutbursts
                    and $59$ normal outbursts.
  The vertical red lines indicate
the local maxima for the normal outbursts 
  in which coverage permits a reliable determination,
    and with well-sampled decays
  down 
   to $\sim$2 mag below maximum.
\label{fig1}}
\smallskip
\end{figure*}

There are two bright SU UMa stars in the {\it Kepler} field
  exhibiting a variety of temporal behavior that make 
  them worthy of a detailed statistical study of their outbursts,
  V1504 Cyg ({\it Kepler} ID 7446357;
  $P_h=1.67$) 
 and V344 Lyr ({\it Kepler} ID 7659570;
  $P_h=2.1$). These are members 
          of the VW Hyi subdivision.
  To date the two light curves have amassed 736.4 d at
 1$-$min cadence. Excluding gaps and bad data points, 
 the light curves contain 1000431 and 1000345 data entries,
  respectively. 
Previous studies  of the  {\it Kepler} data  on SU UMa stars
   have  found  
quiescent superhumps in V344 Lyr
   (Still et al. 2010),
    presented numerical models 
    of the long term light
    curve of  V344 Lyr (Cannizzo et al. 
   2010; hereafter C10),
    and studied superhumps, both
 positive and negative,  in  the long term V344 Lyr 
    light curve  (Wood et al.  2011).

\begin{figure*}
\centering
\epsscale{1.0}
\includegraphics[scale=0.75]{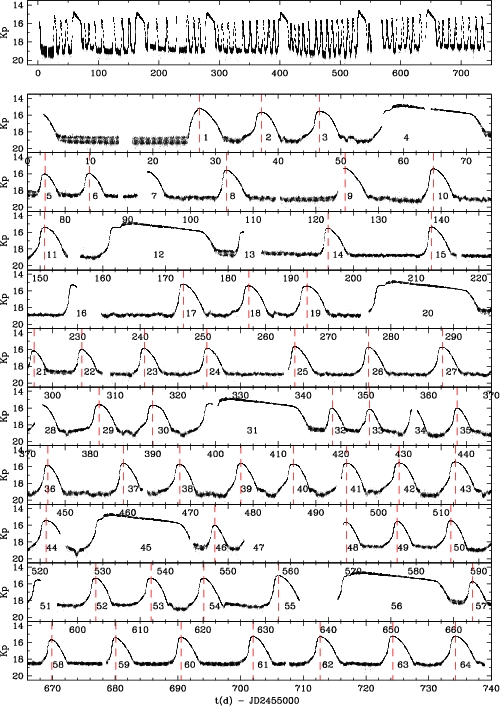}
\vskip .35cm
\figcaption{
       {\it Kepler} light curve
   for  V344 Lyr at 1 min cadence, spanning $\sim$736 d.
   The light curve encompasses six  superoutbursts
                and $58$ normal outbursts.
  The conventions are the same as in Figure 1.
\label{fig2}}
\smallskip
\end{figure*}

Statistical studies of DNe have been useful in delineating 
  the   long-term properties of outbursts, e.g.., the outburst
duration, recurrence times, and quiescent intervals,
   and placing physical constraints on models of outbursts
(e.g., 
          Campbell  \& Shapley 1940,
 Sterne,  Campbell  \& Shapley 1940,
          Bath \& van Paradijs 1983,
                  van Paradijs 1983,
             Szkody  \& Mattei 1984,
            Cannizzo \& Mattei 1992, 
                               1998,
                    Ak, et al. 2002,
         Simon 2004).
   Several interesting studies of the SU UMa stars have been
   carried out. 
  For instance, van Paradijs 
    (1983)
  studied the variation
  in outburst duration $t_b$ 
   with orbital period
   for 15 DNe 
  spanning the $2-3$ hr CV period gap.
  He  found that 
 short outburst
   durations
   increase with orbital period,
   whereas 
long outburst
  durations  
   are relatively constant 
  with orbital period.
   Therefore for the SU UMa systems, 
   which lie at short orbital period,
             the ratio  
    $t_b({\rm SO})/t_b({\rm NO})$ is large.
 The relation of superoutbursts to normal outbursts
for DNe below the period gap and the 
relation of long outbursts to normal outbursts 
  for DNe above the period gap are part of a general trend;
  superoutbursts are just long outbursts
  in 
   short orbital period DNe.
   This finding was amplified by Ak et al. (2002)
using a larger sample.
   In this work we seek to extend 
  previous studies by analyzing the high fidelity
{\it Kepler} light curves of two SU UMa stars
to investigate the properties of the outbursts.
  In Section 2 we examine the outburst 
  properties of the  NOs and SOs,
  in  Section 3 we  discuss the results, in particular
  the scaling of the SO decay rate with orbital period,
  and
  in Section 4 we summarize our results.

\section{Outburst Properties}

\subsection{Normal Outbursts}

The relatively small number
  of total  outbursts observed 
  in the two systems to date does not
  yet allow one to amass detailed
   frequency  histogram  distributions,
  as was done for instance by Cannizzo 
 \& Mattei (1992) for the $\sim$700
outbursts observed up to that time in SS Cyg.
   However,
  the variations of the outburst duration $t_b$ and 
intervening quiescent interval $t_q$ within a supercycle
contain information which can potentially guide
  theoretical models.  
   For instance, in the thermal-tidal
   model  for  the supercycle in the SU UMa systems
  (Ichikawa \& Osaki 1992, 1994;
   Ichikawa, Hirose, \& Osaki 1993),
  one sees a monotonically increasing sequence
 of NO quiescent intervals leading up to a SO.
    Now with the exquisite {\it Kepler}
   data,  this can be directly
   tested. 
   The NOs of the fainter SU UMa stars
  such as V344 Lyr have 
     been poorly characterized prior to {\it Kepler} 
       due to their intrinsic faintness
  (see, e.g., 
      Fig. 1 of Kato et al. 2002).

Figures 1 and 2 present the currently available 
  long term light curves for
    V1504 Cyg and V344 Lyr.
   To obtain the apparent {\it Kepler} magnitude
  from the original data, electron flux per cadence,
   a correspondence\footnote{See
   http://keplergo.arc.nasa.gov/CalibrationSN.shtml.}
   of $Kp=12$ to $10^7$ $e^{-}$ cadence$^{-1}$
  was adopted. 
  Each light curve encompasses six SOs;
  V1504 Cyg had 59 NOs while V344 Lyr had 58.
  Two systematic effects
  are worth noting:
(i) There are   gaps in the data 
      due to monthly data downloads,
   quarterly $90^{\circ}$ spacecraft rolls,
  and spacecraft safe modes.
(ii) In addition,  there
  is no
    long-term
absolute flux calibration for {\it Kepler}
     (Haas et al. 2010,
      Koch et al. 2010):
(1)  Target masks do not contain all the flux from the target. 
 The aperture is chosen 
 to maximize a varying S/N over the whole quarter.
As {\it Kepler} orbits the sun,
 however, targets drift
across the mask as a result 
   of differential velocity aberration. 
Varying amounts of target flux fall
outside of the pixel aperture over the quarter, 
  so the 
 target flux tends to vary smoothly over the quarter due to
this effect.
   Another contributing factor
   to this effect
   is focus variation due to the
   thermal behavior following spacecraft slews,   
   especially immediately after Earth 
   pointings for data download,  or safe modes.
(2)
    After each {\it Kepler} rotation, the target falls
    on a different CCD chip with a different PSF and 
   the target is centered
 differently, e.g., perhaps on a pixel boundary 
  rather than a pixel center. The 4 arcsec pixels undersample the
PSF. This means that the pixel mask changes shape and
   size each quarter, with different amount of light losses.
(3)
 With different shaped masks, the number and amount of contaminating sources in the mask will differ across
the quarter gap.

 The lack of 
    long term flux
    calibration is evident
    in the elevated flux level seen in V1504 Cyg 
    during $100 \la
   t({\rm d}) \la 200$ (i.e, Q2 of observations).
     In addition the secular trending of the
     flux calibration within a given quarter
     may account for the apparent
           $\sim$50 d $e-$folding decrease
  of the quiescent level
   following the first two SOs in V344 Lyr,
  both of which occurred just after a roll maneuver.
 In view of these considerations,
  we limit our attention in this study
  to timescales rather than amplitudes.

\begin{figure}
\centering
\epsscale{1.0}
\includegraphics[scale=0.5]{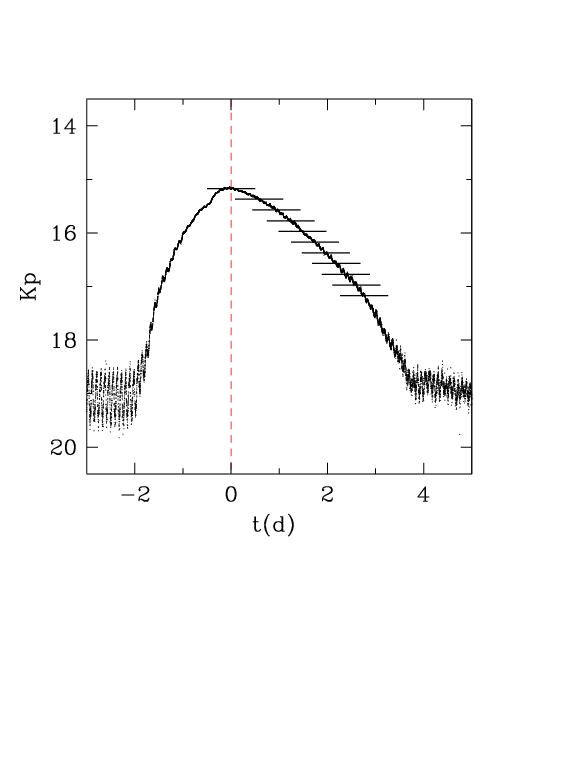}
\vskip -3.5cm
\figcaption{
The first 
  normal outburst in V344 Lyr,
with
   eleven horizontal line
segments along the decay 
 at equally spaced
   intervals of $\Delta K_p=0.15$.
  The segments define the edge of 
  the ten bins used
in calculating the histogram of decays
shown in Figure 4.   
   The decay for each NO is  
   divided into  similar bins,
  using the local maximum of each NO as 
the starting point.
 The local decay rate within a  bin
is defined to be $t_d \equiv 
 (t_j-t_{j-1})/(Kp_j-Kp_{j-1})$,
  where the index $j$ denotes a bin edge,
  $1 \le j \le 10$.
\label{fig3}}
\smallskip
\end{figure}

In a standard light curve plot of magnitude versus time,
 a straight line decay means 
  that the flux $f =f_0 e^{-t/\tau}$,
which we would therefore refer to as 
   an exponential decay.
   A decay which deviates from exponential
  could  
be concave   upward  (slower-than-exponential),
   concave downward  (faster-than-exponential),
   or some more complicated shape. 
  The  decays of the NOs are 
       faster-than-exponential,
   which obviates theoretical  
   studies that attempt   to introduce
a weak radial power law
   dependency  into the SS $\alpha$ parameter
   in  order to get precisely exponential 
   decays 
  (e.g., 
    Ichikawa, Hirose, \& Osaki 1993;
                      Cannizzo 1994).
We may quantify the deviation from
exponentiality in the NO decays
 by dividing the decay into bins,
each spanning a small range $\Delta Kp=0.15-0.20$,
  and  calculating the locally 
 defined decay rate 
      $t_d$(d mag$^{-1}$)
      within each bin.
Dividing the decay into a total of 10 bins for each star,
we take $\Delta Kp=0.15$ for V1504 Cyg and 
        $\Delta Kp=0.20$ for V344 Lyr.
     For demonstration purposes, the binning distribution
  for the decay of the first NO in  V344 Lyr
  is shown in Figure 3.
 Within each bin we compute a frequency histogram
  distribution of $t_d$ values for all  NO decays.
   The results are shown in 
   Figures 4 and 5.    
   The small numbers in each panel indicate
  the magnitude range in $Kp$ measured from the
 local maximum for each NO.
   Thus, 
  the upper panels show the decay
rates computed near maximum light, and as  one progresses
to lower panels one is sampling the fainter portions.
For both systems one sees a shifting in the centroid
of the frequency histogram 
      distribution to lower $t_d$ values as
the decay proceeds, equivalent
to saying that the decays are concave downward.
  In both systems, 
    the parts of the decays
  fainter than about 1 mag  below NO maximum
  asymptote to  a centroid median 
     of $\sim$0.6$-$0.7 d mag$^{-1}$.
  Thus the concave-downward aspect of the 
  NO decays
   is due  in large part
   to the rounded
 maxima; by the time the NO has faded
   to within $\Delta Kp \approx 1 $ mag
   of quiescence,
the  rest of the decay is closely exponential.

\begin{figure}
\centering
\epsscale{1.0}
\includegraphics[scale=0.5]{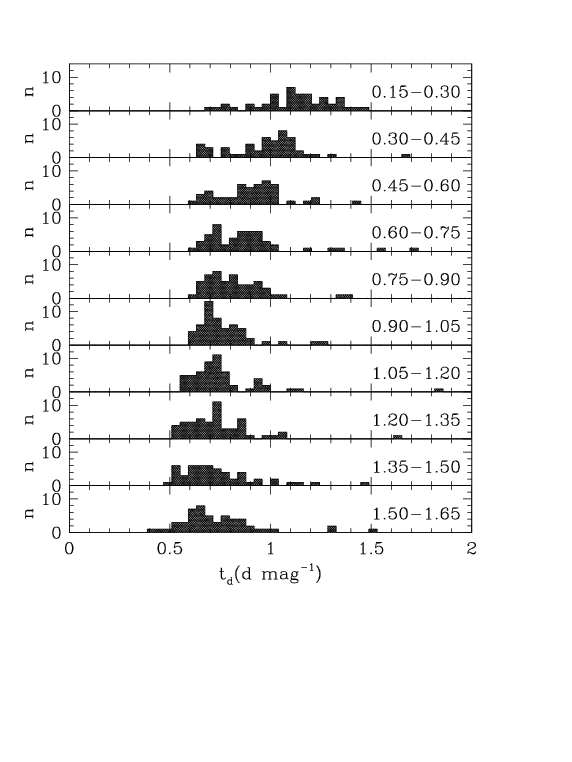}
\vskip -2.75cm
\figcaption{
  A histogram of the fast decay rates
  for the 54 normal outbursts in V1504 Cyg
  with well-defined maxima
  and well-sampled decays,
  as 
  shown in Figure 1.
  The numbers in each panel give  the 
  magnitude range in $\Delta Kp$ 
down  from maximum
corresponding to the interval within which the
local decay rate $t_d$ is measured,
  thus brighter portions of the decay 
are in the upper panels.
\label{fig4}}
\smallskip
\end{figure}

\begin{figure}
\centering
\epsscale{1.0}
\includegraphics[scale=0.5]{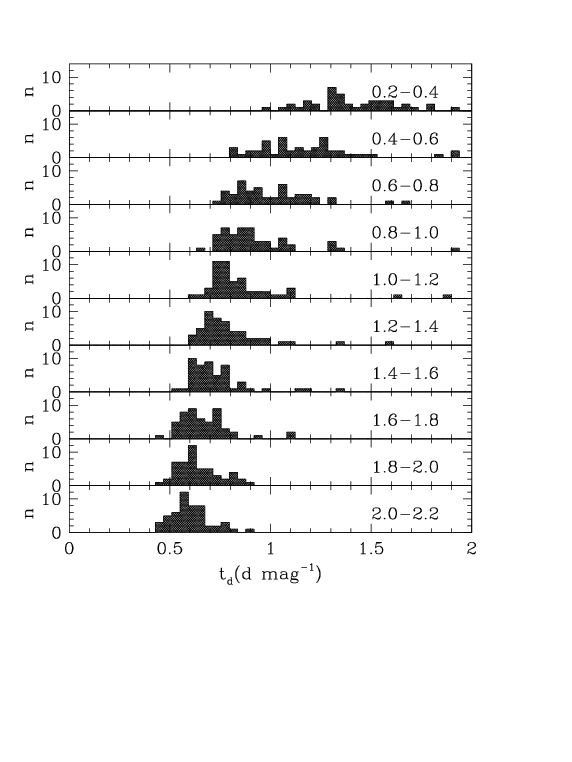}
\vskip -2.75cm
\figcaption{
  A histogram of the fast decay rates
  for the 51 normal outbursts in V344 Lyr
    with well-defined maxima
    and well-sampled decays, as shown in Figure 2.
    Conventions are the same as in Figure 4.
\label{fig5}}
\smallskip
\end{figure}

  Bailey (1975) 
   discovered that the  decay rate for NOs
is correlated with orbital period 
   (see his Figure 1 for eight well-studied
systems), and derived the empirical relation 
$t_d = 0.38$ d mag$^{-1}$ $P_h$.
  Smak (1984)
  was the first to show that
  this relation
   can be used to infer $\alpha_{\rm hot}\simeq0.1$.
      Subsequent
   work confirmed and expanded on this finding 
  (Cannizzo 2001a).
    Warner (1995b)
  looked at a larger sample than Bailey
    and
  deduced the fit
  $t_d = 0.53$ 
 d mag$^{-1}$  $P_h^{0.84}$.
  For the two systems of interest in this study,
  for which $P_h=1.67$  (V1504 Cyg)
                and 2.1 (V344 Lyr), respectively,
the predictions for the rates of decay would be
0.64 and 0.80 d mag$^{-1}$ from the original Bailey relation,
 and 0.82 and 0.99 d mag$^{-1}$ from the Warner fit.
  The presently determined values $\sim$0.7 d mag$^{-1}$ 
  in V1504 Cyg and  $\sim$0.6 d mag$^{-1}$ in V344 Lyr
  for the faintest bins
 are closer  to Bailey's original expression.
    If one averaged over the entire decay, 
   the values would be larger,
  closer to Warner's expression.
The precision of the current determinations  are limited 
 by the small number of total 
NOs in both systems. With time, better statistics will allow 
 better determinations, as has been done for instance
with SS Cyg (Cannizzo \& Mattei
      1998, 
    see their Figure 6).

Figure 6 shows the outburst durations
    $t_b$
  for the two systems. 
   Based on the magnitude range for outbursts
  in the two systems, 
       we use a cut of $Kp=16$   in V1504 Cyg 
                  and  $Kp=17.9$ in V344 Lyr
   to define outburst onset and termination.
   There is an obvious 
   trend
   of 
 increasing outburst width  $t_b$
  between consecutive
  SOs:  during a supercycle,
   $t_b$ increases by
   $\sim$20$-$70\%. 
  Figure 7 shows the asymmetry parameter $f_r$,
  defined to be the ratio of the  rise time 
 to the total outburst duration,
for the individual NOs depicted in Figures 1 and 2.
Figure 8 shows the quiescent intervals
  $t_q$
  between successive outbursts.
 In both systems there is a general
  trend of
  $t_q$  values
  that increase to a local maximum
about mid-way between successive SOs.
    For the last half of the V1504 Cyg light 
curve  and  the first half of the V344 Lyr
   light curve this effect is quite pronounced.
 In is interesting that
  while the values of $f_r$ and
  $t_q$ from one NO to the next are fairly
  continuous, even across an SO, 
  the $t_b$ values
   show a sharp discontinuity across an SO, resetting
to a much smaller value.

At roughly the time of the third SO in V1504 Cyg,
    strong
negative superhumps appear (Wood et al. 2011, in preparation)
  and  persist  through the end of the light curve.
 The  presence of negative superhumps indicates
 that the mass is not accreting primarily at the rim of the disk.
   Although there is not an obvious change in $f_r$
  at this time, it is perhaps noteworthy that $t_q$
 increases. This may indicate a strong drop in
 the mass transfer rate from the secondary star, which would
increase the recurrence time for outbursts.
    In the magnetic systems such as AM Her, which
lack accretion disks,  one has direct 
evidence for the instantaneous mass transfer
   rate from the secondary stars in the long
term light curves, and one does often
see intervals during which the  mass transfer rate
reduces to a low level 
 (e.g., Hessman, G\"ansicke, 
 \& Mattei 
 2000).





\subsection{Superoutbursts}

Figure 9 shows a composite of the six  SOs
  in V1504 Cyg, positioned in time so that
$t=4$ d corresponds to the crossing of $Kp=16$
  during SO onset.
  The rate of decay during SO is $\sim$8 d mag$^{-1}$.
Figure 10 shows a composite of the six SOs
  in V344 Lyr, positioned in time so that
$t=3$ d corresponds to the crossing of $Kp=18$
  during SO onset.
  The rate of decay during SO is $\sim$12 d mag$^{-1}$.
 Within each system  the SOs  are quite similar  in overall
appearance and duration. 

\section{Discussion}

  A summary of the parameters associated with the outbursts
 in the two systems is given in Table 1.
The behavior of  the sequences of
    outburst durations $t_b$ and
   quiescent intervals $t_q$ during a supercycle
   are
  quite different.
The $t_b$ sequence shows an increase
   between SOs in both systems. This is consistent
with a secular increase in the mass of the accretion disk.
   The  $t_q$ values, by contrast, 
   increase to
  a local maximum roughly half way between SOs
   before returning to their initial small
  value by the time the next SO begins. Thus the
$t_q$ values are roughly continuous across SOs.
In both systems
  the  variation of $t_q$ values between SOs appears to be 
   inconsistent with the prediction
  of the thermal-tidal instability  model,
  which shows a monotonic increase in $t_q$
   between SOs 
 (Ichikawa, Hirose, \& Osaki 1993, see their Figure 1;
                       Osaki 2005,  see  his Figure 3).
  This indicates that the triggering radius for NOs 
does not move uniformly outward with each successive NO,
but rather attains a local maximum and then recedes.

\begin{figure}
\centering
\epsscale{1.0}
\includegraphics[scale=0.5]{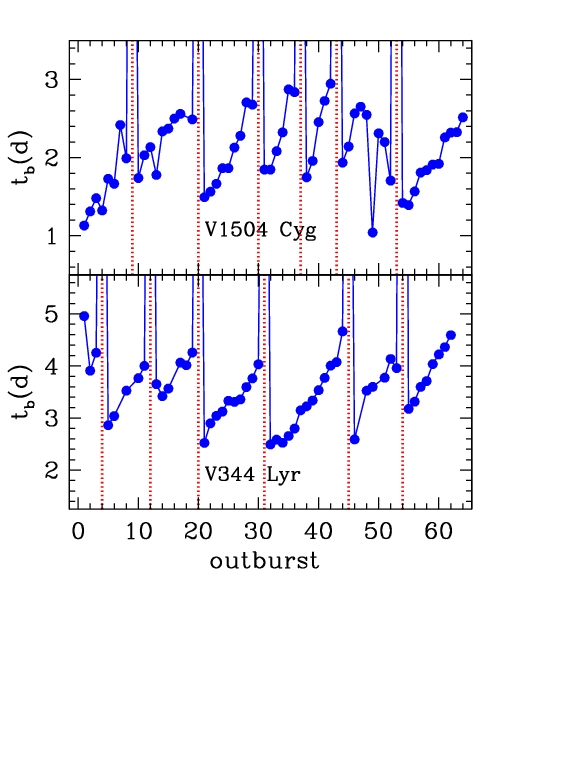}
\vskip -2.75cm
\figcaption{
The values of the outburst
  durations $t_b$ for the two systems,
   taking a cut line $Kp_{\rm cut}=16$
   for V1504 Cyg and $Kp_{\rm cut}=17.9$
   for V344 Lyr. 
  The vertical dotted lines ({\it red})
  indicate the superoutbursts.
  The values of $t_b$ for the superoutbursts
  are $t_b$(d) = 12.1, 12.0, 11.9,  12.0, and 12.0
  for V1504 Cyg, and  
      $t_b$(d) = 16.7, 17.1, 16.7,  16.9, and 16.0
   for V344 Lyr. Only five SOs are quoted for each system
  because of a data gap at  the start of the sixth SO in 
  both systems.
\label{fig6}}
\smallskip
\end{figure}

\begin{figure}
\centering
\epsscale{1.0}
\includegraphics[scale=0.5]{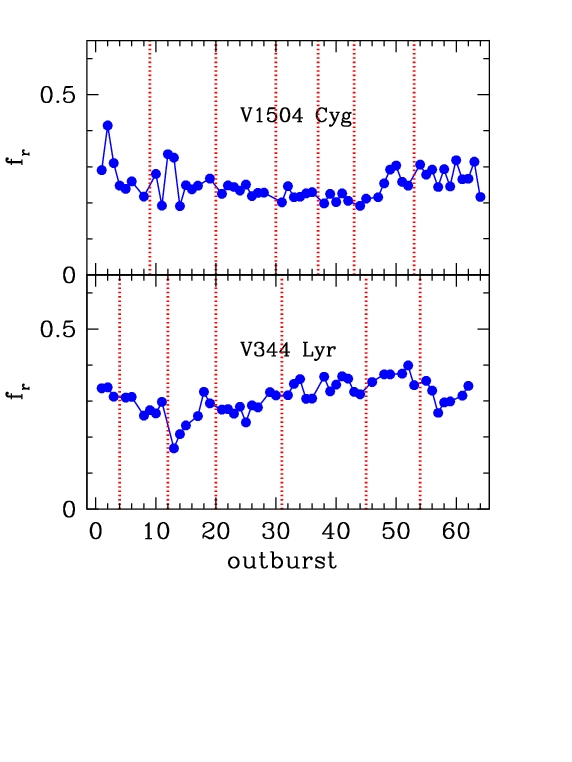}
\vskip -2.75cm
\figcaption{
The values of the 
   asymmetry parameter $f_r\equiv t_{\rm rise}/t_b$
 for the normal outbursts in the two systems.
 Only values for which the
  times of outburst
  maxima can be reliably determined
   are shown.
   The same cut line values are used as in Figure 6.
\label{fig7}}
\smallskip
\end{figure}

%
\begin{deluxetable}{cccc}
\tablewidth{0pc}
\tablenum{1}
\tablecaption{Summary of Parameters}
\tablehead{
\colhead{Parameter}   &
\colhead{V1504 Cyg}   &
\colhead{V344 Lyr}    &
\colhead{Comment}
}
\startdata
NO: $t_b$(d)             & 1.1$-$2.9    & 2.5$-$5.0    & NO duration \\
SO: $t_b$(d)             &  $12\pm0.07$ & $16.7\pm0.4$ & SO duration \\
NO: $f_r$                & 0.19$-$0.41   & 0.17$-$0.40   & NO rise/duration \\
    $t_q$(d)             & 3.2$-$20.4    & 2.5$-$21.5  & Quiescent duration  \\
NO: $t_d$(d mag$^{-1}$)  &       0.7     &     0.6     & NO decay rate   \\
SO: $t_d$(d mag$^{-1}$)  & $8.1\pm0.4$   &  $12\pm0.6$ & SO decay rate
\enddata
\end{deluxetable}

 The asymmetry of the NOs is similar in the two systems,
$f_r\simeq 0.2-0.4$.
 According to theory, the degree of outburst asymmetry
 reveals whether the outburst is triggered at small or large 
radii in the disk (Smak 1984; Cannizzo, 
 Wheeler, \& Polidan 1986, hereafter CWP;
  Cannizzo 1998,
    see his Figure 6 and associated discussion). 
Outbursts triggered at large radii tend to have fast,
  asymmetric
rise times ($f_r< 0.5$), while those triggered at small radii
 are more symmetric ($f_r\simeq 0.5$).
 Smak referred to these two types as Type A and Type B, respectively,
whereas CWP used the terminology
     ``outside-in'' and ``inside-out'' outbursts.
 The low $f_r$ values in both systems studied in this work are consistent
  with the Type A/outside-in outbursts.
  It is interesting that the one outburst which seems to be
  certifiably symmetric and therefore presumably inside-out
   $-$ 
  outburst 1 of V344 Lyr
   $-$ 
  occurred after an unusually  long quiescence
interval, which may have led to a more filled disk at the end of 
  quiescence (compare this outburst with that shown in
        Figure 5b of Smak 1984   
         \ 
   and  Figure 3 of CWP).
 During this part of the  V344 Lyr light curve
  there were strong negative superhumps (Wood et al. 2011, see
their Figure 11), indicating a tilted accretion disk. This would
have the effect of spreading the accretion stream impact
  over a large range in radii, thereby lengthening
   the time to build up  to the next outburst.

\begin{figure}
\centering
\epsscale{1.0}
\includegraphics[scale=0.5]{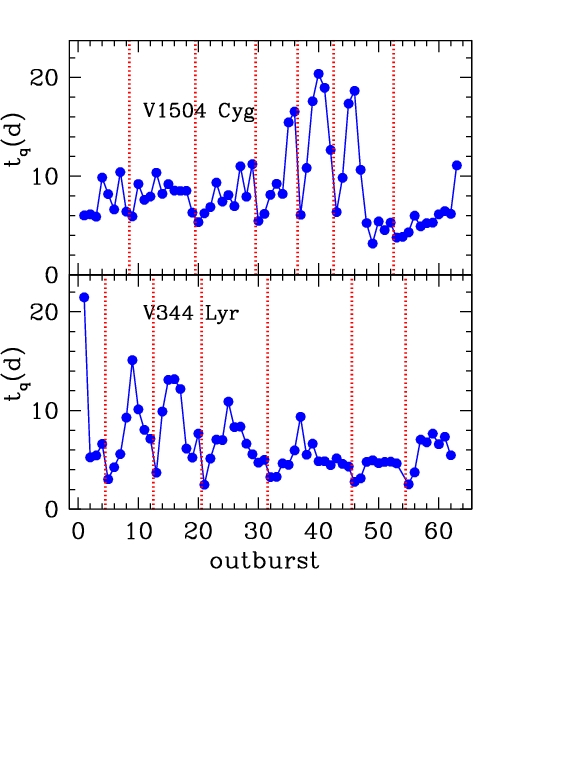}
\vskip -2.75cm
\figcaption{
The values of the quiescence
  durations $t_q$ between consecutive 
   outbursts for the two systems.
As with Figure 6, the vertical dotted lines ({\it red})
  indicate the positions of the superoutbursts.
   The same cut line values are used as in Figure 6.
\label{fig8}}
\smallskip
\end{figure}

 The mean durations of SOs for V1504 Cyg and V344 Lyr are
11.9 d and 16.8 d, and the $e-$folding rates
    of the (slow) decay during SO 
   are  $8.1\pm0.4$ d mag$^{-1}$ and $12.0\pm0.6$ mag$^{-1}$,
   respectively, where the errors indicate the standard
deviations of the six measurements.
 Thus the SOs  span $\sim$1.5 $e-$foldings in V1504 Cyg,
and $\sim$1.4 $e-$foldings in V344 Lyr.
          Given that the SO accretion disks are close to 
steady state,  the depletion fraction $f_d$ in the stored disk
   mass during SO for the two systems,
\begin{equation}
    f_d \simeq  1-e^{-t_b({\rm SO})/t_d({\rm SO})},
\end{equation}
  is $\sim$0.78 and  $\sim$0.75, respectively. 
(These numbers probably over-estimate the depleted mass
   fraction because they encompass portions of the SO
 fainter than the $e-$folding plateaus.) 
  These depletion fractions are larger
    than the $f_d\simeq0.2$ inferred
  for long outbursts in DNe above the period gap (e.g., 
      Cannizzo 1993),
  and stem  from the larger 
    $\alpha_{\rm hot}/\alpha_{\rm cold}$ ratios in  the SU UMa systems.
  The ratio $t_d({\rm SO})/t_d({\rm NO}) \simeq 20$ is consistent
with the previous interpretation of fast decays being due to the action
of a cooling front within the disk, and slow decays being the result
of gradual accretion onto the WD, and therefore governed by the viscous
time scale in the outer disk. 
   The viscous and thermal time scales can be
written in terms of the local Keplerian rotation frequency
  $\Omega$ and the local disk height to radius
  ratio $h/r$ (Pringle 1981).
     The viscous time scale  
\begin{equation}
 t_{\nu} = 
{1\over {\alpha_{\rm hot}\Omega} } \left(r\over h\right)^2
\label{visc0}
\end{equation} 
  and the thermal time scale  
\begin{equation}
t_{\rm th} =
{1\over {\alpha_{\rm hot}\Omega} } \left(r\over h\right)
\end{equation} 
   differ by a factor $h/r$,
  therefore $ h/r \approx  1/20$.

C10 present   an equation for the rate of
decay of a SO, taken from 
    Warner (1995a, 
            1995b),
  which  is based on the viscous time 
  in the outer accretion disk
\begin{equation} t_d 
  \approx  
 t_\nu =  17 \ {\rm d} 
               \ \alpha_{{\rm hot},-1}^{-4/5}
               \ P_h^{1/4}
               \ m_1^{1/6},
\label{brian}
\end{equation}
  where $\alpha_{{\rm hot},-1}=\alpha_{\rm hot}/0.1$
   and 
  $m_1=M_1/\msun$. 
  This $e-$folding time scale
  should approximate the
  decay rate, expressed in d mag$^{-1}$.
 Hence in a large sample of DNe
 the prediction is that the rate of
decay during SO should vary as $P_h^{0.25}$.
Our analysis shows, however, that in going from 
   V1504 Cyg with $P_h=1.67$ 
 to V344 Lyr with $P_h=2.1$, the SO decay
rate increases from 8 d to 12 d.
  One does expect object-to-object variations
in NO and SO properties which can be large
  compared to overall, secular trends with
orbital period, but given the dependencies
 in Equation (\ref{brian}) and the inference from
 a large number of DNe that $\alpha_{\rm hot}\simeq0.1$,
   there is no way to account for such
a steep variation, 
   $t_d \simpropto P_h^{1.8}$.

\begin{deluxetable}{cccc}
\tablewidth{0pc}
\tablenum{2}
\tablecaption{Slow Decay of Superoutbursts in SU UMa stars}
\tablehead{
\colhead{System}       &
\colhead{decay rate
     (d mag$^{-1}$)}   &
\colhead{$P_h$}        &
\colhead{Reference}
}
\startdata
SW UMa    & 10.   & 1.36 & 1 \\
WX Cet    & 10.   & 1.40 & 2 \\
FL Tra    &  7.7  & 1.41 & 3 \\
V1028 Cyg &  8.3  & 1.45 & 4 \\
HO Del    &  7.1  & 1.50 & 5 \\
GO Com    &  6.7  & 1.58 & 6 \\
NSV 4838  & 10.   & 1.63 & 7 \\
VW CrB    & 10.3  & 1.70 & 8 \\
EG  Aqr   &  7.1  & 1.83 & 9 \\
V503 Cyg  &  9.1  & 1.86 & 10
\enddata
\tablerefs{
   1. Soejima et al. (2009),
   2. Kato et al.    (2001),
   3. Imada et al.   (2008b),
   4. Baba et al.    (2000),
   5. Kato et al.    (2003),
   6. Imada et al.   (2005),
   7. Imada et al.   (2009),
   8. Nogami et al.  (2004),
   9. Imada et al.   (2008a),
  10. Corrado et al. (2000)}
\end{deluxetable}

It is worth pausing for a moment
   and looking at 
    how
   the 
   SO slow decay rates for  V1504 Cyg  
     and V344 Lyr 
     fit into the larger  picture.
 Table 2  lists the slow
decay rate of SOs for SU UMa stars taken from the
literature,  most of which are derived from VSNET data 
(see Kato et al. 2009,
                 2010).
             We only show
  data for the VW Hyi subclass 
       for which the
 slow decay rate is relatively constant over the course of the SO.
 In Figure 11 we plot the SO decay
rates for the systems shown in Table 2,
  as well as V1504 Cyg, V344 Lyr, 
 WZ Sge and U Gem. 
   For WZ Sge we only plot the value
for the initial, steep rate of decay, 4 d mag$^{-1}$.
As noted in the Introduction,
   the SO decays in the  WZ Sge stars  
   exhibit a strong deviation from
   exponentiality
          presumably
due to the accretion of virtually all  the
stored mass. As the disk mass decreases,
   one expects a concomitant increase  in
the viscous time,
    leading to a pronounced concave upward SO decay.
    Therefore only the very initial SO decay
rate gives a valid representation for the viscous time 
 within the disk corresponding to the stored mass. 
    As regards DNe above the period gap, 
  virtually none
  of the
long outbursts ever observed exhibited enough
dynamic range in their overall decay that one could  reliably
extract the rate of decay. Only for the unusually long, 1985 October
 outburst of U Gem could this determination be made (Cannizzo, Gehrels,
and Mattei 2002),
   hence the single data point above 
   the period gap
  in Figure 11.
  If one takes the overall $t_d - P_h$
  trend between WZ Sge and U Gem as 
 a true physical baseline,
   $t_d \simpropto P_h^{1.6}$,
   the 
 trend between
V1504 Cyg and  V344 Lyr
        in $\log t_d - \log P_h$ space,
    $t_d \simpropto P_h^{1.8}$,
   is roughly the same, although the normalization
  is offset slightly. 
  Given the vertical scatter in Figure 11 for the other SU UMa
  systems, this coincidence is fortuitous.

\begin{figure}
\centering
\epsscale{1.0}
\includegraphics[scale=0.5]{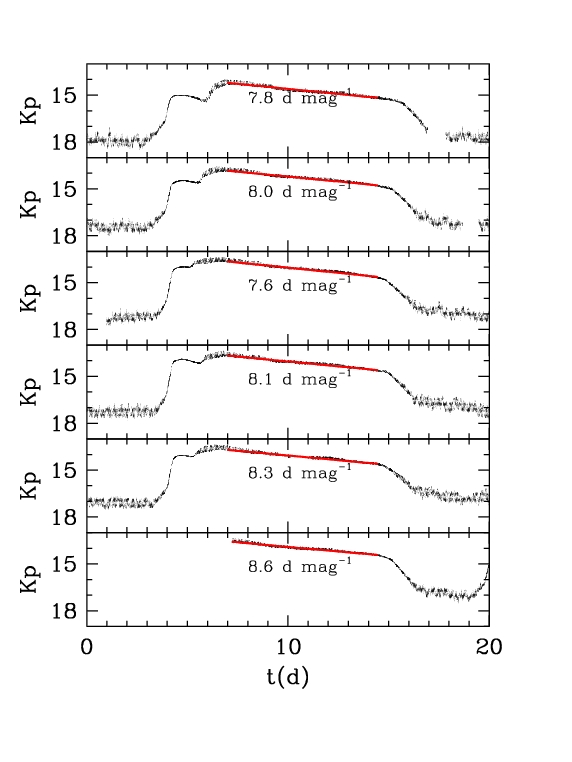}
\vskip -1cm
\figcaption{
A composite of six  superoutbursts
  in V1504 Cyg, positioned in time so that
$t=4$ d corresponds to the crossing of $Kp=16$
  during superoutburst onset.
 The line segments ({\it red})
  indicate the linear least squares fit,
  for the range of data indicated.
The rate of decay for the six  superoutbursts
  are as shown in each panel.
  The mean value is 8.1 d mag$^{-1}$.
\label{fig9}}
\smallskip
\end{figure}

\begin{figure}
\centering
\epsscale{1.0}
\includegraphics[scale=0.5]{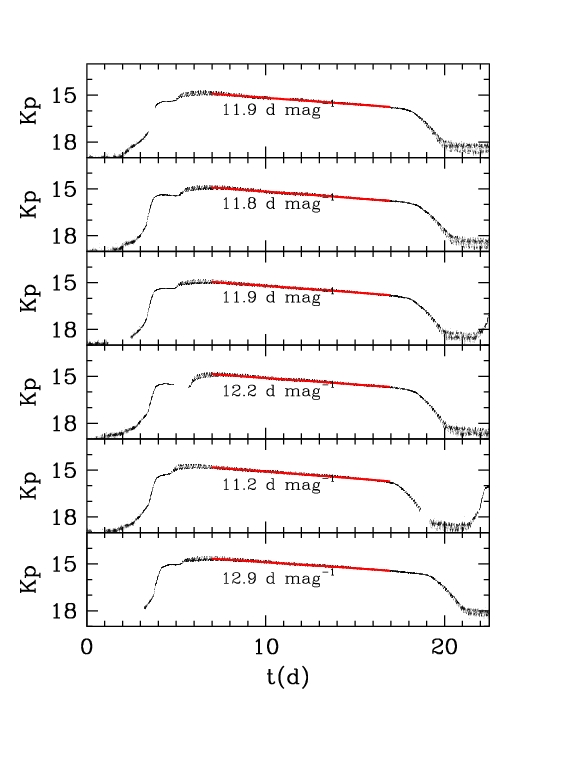}
\vskip -1cm
\figcaption{
A composite of six  superoutbursts
  in V344 Lyr, positioned in time so that
$t=3$ d corresponds to the crossing of $Kp=18$
  during superoutburst onset.
 The line segments ({\it red})
  indicate the linear least squares fit,
  for the range of data indicated.
The rate of decay for the six  superoutbursts
  are as shown in each panel.
  The mean value is 12.0 d mag$^{-1}$.
\label{fig10}}
\smallskip
\end{figure}

  The failure of Equation (\ref{brian})
 to account for the difference between
  the
SO slow decay rate in V1504 Cyg and V344 Lyr
indicates that a more sophisticated
  expression may be required.   One obvious deficiency
is that there is no dependence on $\alpha_{\rm cold}$
     which controls
the available amount of fuel in quiescence to power a SO.
The physics behind the viscosity in quiescence  is unknown 
at present, and one could in principle have large 
variations in  $\alpha_{\rm cold}$ between systems at 
     similar orbital periods.
  Let us go back to the definition of the 
  viscous time scale 
  $t_\nu = (\alpha_{\rm hot} \Omega)^{-1}
    (r /h)^2$.
    Since the controlling time scale in this situation
  is the slowest one, this expression is to be evaluated
at the outer disk radius $r_d$. 
 Following Cannizzo (2001b),  we consider the hot
  state of the disk by adopting a Rosseland mean
opacity $\kappa = 2.8 \times 10^{24}$
 g cm$^{-2} \rho T^{-3.5}$
 and mean molecular weight $\mu=0.67$.
  We can then write the viscous time as
\begin{equation} 
  t_\nu = 5.7 \ {\rm d}  \ m_1^{5/14} r_{d,10}^{13/14}
    \alpha_{{\rm hot}, -1}^{-8/7} 
   \left(\Sigma(r_d) \over
    2000 \ {\rm g} \
  {\rm cm}^{-2} \right)^{-3/7},
\label{visc}
\end{equation}
  where $m_1=M_{\rm WD}/1\msun$, 
  $r_{d,10}=r_d/(10^{10} \ {\rm cm})$,
    and  $\Sigma(r_d)$ is the
  surface density at the outer
  disk edge.

Looking at this expression, we see that the 
 Warner (1995a,
         1995b)
equation already contained an implicit assumption about $\Sigma$,
as it only contains 
   $\alpha_{\rm hot}$ and $P_h$.
 To derive a more meaningful expression for
 $t_d \approx t_\nu$  
  in the context of SOs we need to take into account two 
additional pieces of information, namely that
  during a SO essentially all the mass accumulated
  during quiescence is accreted onto the WD, and thereby
lost from the disk, and also that the accumulated mass
  depends on the critical surface density $\Sigma_{\rm max}$,
  which depends in turn on $\alpha_{\rm cold}$.
  This scaling is given by (Lasota 2001)
\begin{equation} 
  \Sigma_{\rm max} = 13.4 \ {\rm g} \  {\rm cm}^{-2}  \  m_1^{-0.38} r_{10}^{1.14}
    \alpha_{\rm cold}^{-0.83}.
\label{lasota}
\end{equation}
   Since the attainment of $\Sigma_{\rm max}$
  is the triggering mechanism for a DN 
 outburst, the quiescence disk mass 
    $\sim$$(\pi/2) r_d^2 \Sigma_{\rm max}(r_d, \ \alpha_{\rm cold})$
  at the time of SO onset
can be written as
\begin{equation} 
 \Delta M_{\rm cold}  \approx 
     4.2 \times 10^{21} \ {\rm g} \  m_1^{-0.38} r_{d,10}^{3.14}
    \alpha_{\rm cold}^{-0.83}.
\label{mass}
\end{equation}

Once the SO has started, the mass that was present in the non-steady
 state quiescent disk gets redistributed into a quasi-steady
disk with ${\dot M}(r) \simeq $ constant, and therefore with a radial
  profile $\Sigma(r) \simpropto r^{-3/4}$ (SS).
  Therefore  the mass of the hot disk can be written as
\begin{equation} 
 \Delta M_{\rm hot} = \int 2 \pi r dr \Sigma(r, \ \alpha_{\rm hot}).
\label{mass}
\end{equation}
     Since the ratio $\alpha_{\rm hot}/\alpha_{\rm cold}$
is large in SU UMa systems, a substantial amount of mass needs to drain
onto the WD before the surface is depleted enough for the cooling wave
to start.

  We now set $\Sigma(r, \ \alpha_{\rm hot})$ 
  equal to its standard functional 
  form for a SS disk with gas pressure and
  free-free opacity, $\Sigma_{\rm hot} 
    = \Sigma_0 \ \alpha_{\rm hot}^{-4/5} r_{10}^{-3/4}$.
 This is the same form that entered into Equation (\ref{visc}).
  Equating the disk mass
  at the end of the quiescent state
  with that at the beginning of the SO,
    i.e.,  $\Delta M_{\rm cold}
        =
      \Delta M_{\rm hot}$, gives
\begin{equation} 
 \Sigma_{\rm hot}(r_d)  = {5\over 16} \Sigma_{\rm max}(r_d),
\label{mass}
\end{equation}
  the relevant $\Sigma$ value
for  Equation (\ref{visc}).
Therefore we can rewrite Equation (\ref{visc})
as
\begin{equation} 
 t_\nu = 80.2 \ {\rm d} \ m_1^{0.52} \alpha_{\rm cold}^{0.36} 
    \alpha_{{\rm hot},-1}^{-1.14} 
     r_{d,10}^{0.44}.
\label{visc2}
\end{equation}
  This expression differs from Equation (\ref{visc})
  in that it contains $\alpha_{\rm cold}$,  
   reflecting the fact that the total amount of matter
  stored in quiescence enters into the viscous time
  scale in the resultant SO disk.

In extreme mass ratio systems such as the SU UMa stars, the
Roche lobe of the primary asymptotes 
to $\sim$0.82 of the binary separation (Eggleton 1983).
  Also, the relation between orbital separation and orbital period is
  $a_{10} = 3.53 P_h^{2/3}$, where $a_{10}$ is the orbital separation
in units of $10^{10}$ cm.   
   If we further take the outer disk radius $r_d$ to lie at 75\%
  of the primary Roche lobe, we have $r_{d,10} = 2.19  P_h^{2/3}$.
    Since it is fairly certain that $\alpha$ in outburst is always
close to 0.1, $\alpha_{\rm hot}$ been scaled to that value in the previous
equations. On the other hand, the $\alpha$ value in quiescence
  is uncertain, and has been traditionally adjusted  by trial and
error for each system. 
   For SU UMa stars,
     workers
  have typically had to adopt much smaller $\alpha_{\rm cold}$ values
  than for DNe above the period gap.
   Given this uncertainty, let us take $\alpha_{\rm cold}$ to be a power 
   law in orbital period, $\alpha_{\rm cold} = \alpha_{c,0} P_h^{\psi}$.
  Making these substitutions into Equation (\ref{visc2}) we obtain
\begin{equation} 
 t_\nu = 113 \ {\rm d} \ m_1^{0.52} \alpha_{c,0}^{0.36} 
    \alpha_{{\rm hot},-1}^{-1.14}
     P_h^{0.29+0.36\psi}.
\label{visc3}
\end{equation}
The dynamic variation associated with 
   one magnitude (2.512)
  is 0.924 times that associated with
   one $e-$folding (2.718). Therefore if
 we associate  the rate of SO decay,
  expressed in d mag$^{-1}$,
   with the viscous time at the
outer edge, which is an $e-$folding time, then Equation (\ref{visc3})
needs to be corrected by a factor 0.924.

By comparing the SO decay rates of V1504 Cyg and V344 Lyr, 
 we infer a decay rate
    $t_d = 0.924 t_\nu
  = 12 \ {\rm d} \ (P_h/2.1)^{1.8} =
  3.16 \ {\rm d} \ P_h^{1.8}$.
  Furthermore, adopting $m_1=0.6$ for the SU UMa systems gives
  $m_1^{0.52}=0.767$, and
   therefore 
\begin{equation} 
 80.2  \ {\rm d} \ \alpha_{c,0}^{0.36} P_h^{0.29+0.36\psi} 
  = 3.16  \ {\rm d} \ P_h^{1.8}. 
\label{visc4}
\end{equation}
  Solving for the constant and exponent
  in the $\alpha_{\rm cold}$ scaling gives
     $\alpha_{c,0}=1.25\times10^{-4}$
and $\psi =4.2$ $-$  a steep
dependence of $\alpha_{\rm cold}$ on orbital period.
  For  V344 Lyr at $P_h=2.1$ 
  we  obtain $\alpha_{\rm cold}\simeq 2.8\times 10^{-3}$,
  which accounts for why C10, in their modeling of V344 Lyr,
   had to adopt the small value $\alpha_{\rm cold} = 0.0025$.
  The inference $\psi =4.2$
  is only valid, however,
    if the trend between  V1504 Cyg and V344 Lyr
   were representative of a trend with $P_h$,
   which seems unlikely given the data  for the
other SU UMa systems.
  Based only on the SU UMa data,
    it would be more likely that
         $\alpha_{c,0}$ has a large variation
  between the two systems.
  As noted earlier however, 
  if one takes the WZ Sge $\leftrightarrow$ U Gem
   interpolation to be physically meaningful,
      then the  V1504 Cyg $\leftrightarrow$  V344 Lyr
   interpolation 
    is
   fortuitously coincident.

 A  steep dependence of $\alpha_{\rm cold}$  on  $P_h$
  may have observational support from another line of argument.
  In a recent study      of the aggregate
  properties of DNe, Patterson
     (2011,
   see his Figure 11)
  finds that the recurrence time for SOs has a steep inverse
  relation with binary mass ratio
   $t_{\rm recur}({\rm SO}) \propto q^{-2.63\pm0.17}$,
   where $q=M_2/M_1$.
   Since the secondaries in cataclysmic variables are, by 
  definition, in contact with their Roche lobes, this implies that
$M_2$ scales with orbital  period, and hence
   $t_{\rm recur}({\rm SO}) \propto P_h^{-2.63\pm0.17}$.
  Given that (1) SOs accrete virtually all their stored
 mass $\Delta M_{\rm cold}$, (2) the mass feeding rate
  into the outer disk from the secondary ${\dot M}_T$
   is thought to be relatively constant at $\sim$$10^{-10} \msunyr$
  for systems below the period gap (Kolb et al. 1998;
   Howell et al. 2001), 
   and (3) the recurrence time 
    (Cannizzo et al. 1988)
\begin{equation} 
  t_{\rm recur}({\rm SO}) \approx {\Delta M_{\rm cold} \over |{\dot M}_T|}
  \simpropto r_d^{3.14} \alpha_{\rm cold}^{-0.83},
\label{joe}
\end{equation} 
   it is quite surprising that $t_{\rm recur}({\rm SO})$
   apparently has a strong inverse relation with orbital period.
   As one progresses to shorter orbital period, 
   one expects more frequent SOs since the disks
   are smaller and the critical mass is reached more quickly.
   We are seemingly drawn to the same conclusion as previously, 
  that $\alpha_{\rm cold}$ must have a strong dependence on orbital period.
  Howell et al. (2001; see their Figure 2, panel (a))
  indicates a rough dependence $d \log |{\dot M}_T|/d \log P_h
  \approx 1$ for the SU UMa stars. If we again set 
   $\alpha_{\rm cold} \propto P_h^{\psi}$,
   substitute $P_h$ for $r_d$ in Equation (\ref{joe}), 
   and solve for $\psi$, we obtain $\psi=4.5$.

 If the overall trend in $t_d$ for long outbursts
between WZ Sge and U Gem, which is roughly matched (coincidentally)
  by that between V1504 Cyg and V344 Lyr, is real,
   then
    two independent lines of reasoning
     imply a very steep dependence
of $\alpha_{\rm cold}$ on orbital  period. Why should one have
   $\alpha_{\rm hot} \simeq $ constant,  while at the same time
   $\alpha_{\rm cold} \propto P_h^{\psi}$,
  where $\psi\simeq 4.2-4.5$?
       Current thinking on the physical mechanism for viscosity in
accretion disks centers on the  magnetorotational instability (MRI - Balbus
\& Hawley 1998),
    an instability based on the shearing amplification
  of a weak magnetic field.
 The  nonlinear saturation  limit of the  MRI in ionized 
gas is thought to explain why $\alpha_{\rm hot}\rightarrow\sim$0.1 
   (McKinney \& Narayan
       2007a,
       2007b).
    However, as the disk cools to the quiescent state and
the supply of free electrons drops precipitously due to the exponential
dependence of partial ionization on temperature below $\sim$$10^4$ K,
    the  magnetic field no longer couples effectively to the gas, and fails
to provide an effective viscosity.
    Menou (2000)
     tried to model the complete accretion disk limit
  cycle model for DNe using only the MRI, and  failed. 
    He  discussed the failure of the MRI in quiescence, and 
 predicted that if tidal torques
     (a completely different
   physical mechanism from the MRI)
   dominate the angular momentum transport
 in the quiescent accretion disks in the SU UMa stars, one would expect
an anticorrelation between $t_{\rm recur}({\rm SO})$ and   $q$.
  He presents data for six SU UMa stars showing an anticorrelation.
  (Patterson [2011] 
  shows the anticorrelation  in much greater detail.)
  Menou  notes that theories of tidally induced spiral waves 
    or shock waves indicate  a  strong correlation between
effective torque and $q$
  (Papaloizou \& Pringle 1977;
            Goldreich \& Tremaine 1980).
    Our finding $\alpha_{\rm cold} \simpropto P_h^{4.2}$
    seems to support the view that 
       tidal torquing is the 
dominant angular momentum transport mechanism operating
   in quiescent accretion disks in interacting binaries.

For completeness we note that previous workers 
 have commented on the apparent necessity of small $\alpha_{\rm cold}$
to obtain the large amplitude SOs and long recurrence times
in the short orbital period DNe.
  For example Meyer-Hofmeister et al. (1998)
  found that the large required stored disk mass
and outburst long recurrence time in WZ Sge
mandated $\alpha_{\rm cold}\approx 0.001$
in the disk instability model.
  Meyer \& Meyer-Hofmeister (1999) argued
  for
 a possible physical cause 
  in the loss of coronal
activity
in the very cool atmospheres of the low mass K-type 
 secondaries expected in such short orbital period systems,
resulting in lower magnetic fields in the material entrained
in the L1 point
as it feeds the outer accretion disk.
  However, in view of the findings 
      of Menou (2000) showing the lack of MRI in 
quiescent disks, and the fact that if the MRI were operating 
   the strength
of the final field
within the disk would be  unrelated 
   to the amplitude of the seed field, we
find the tidal torquing arguments more persuasive.

   The  $\alpha_{\rm cold}$ value strongly influences
two observables, the decay time of the viscous plateaus, 
and the recurrence times for outburst.
   It is important to verify whether the steep
   $\alpha_{\rm cold}$ relation mandated by the
  WZ Sge $-$ U Gem trend shown in Figure 11 is consistent
with the observed recurrence times for the
two systems.
    In other words, does  the relation 
   $\alpha_{\rm cold} = 1.25\times 10^{-4} P_h^{4.2}$
  applied to both WZ Sge and U Gem
 lead to reasonable recurrence times
  for both systems?
 For WZ Sge, $P_h   = 1.361$ implies
   $\alpha_{\rm cold} = 4.56\times 10^{-4}$.
  From Smak (1993)
  and Steeghs 
et al. (2007)
  we have $m_1 = 1$, $r_{10}=1.1$, and
  ${\dot M}_T=2\times 10^{15}$ g s$^{-1}$.
  (Smak's value for  ${\dot M}_T$ may be too low,
  as it is below the floor value expected solely 
from gravitational radiation [Howell et al. 2001].)
 Using the value for the stored quiescent mass from Equation (7)
  gives $\Delta M_{\rm cold} =  4.37\times 10^{24}$ g.
 Thus the recurrence time for SOs from Equation (13) is
 $\sim66$ yr, versus the observed recurrence time 
  for superoutbursts in WZ Sge (no normal
 outbursts are seen) of $\sim33$ yr.
  For a standard value given by gravitational radiation
  induced mass transfer below the period gap
  ${\dot M}_T \simeq 10^{-10}\msunyr$,
   the recurrence time  would be  $\sim22$ yr.
  For U Gem, 
  $P_h   = 4.183$ implies
   $\alpha_{\rm cold} = 0.051$,
  From Smak (2001)
  we have $m_1 = 1.07$, $m_2=0.39$,
  orbital separation $a=3.53\times 10^{10}$ cm
 $P_h^{2/3}(m_1+m_2)^{1/3}=1.04\times 10^{11}$
cm, and $r_{10}=0.3a_{10} \simeq 3$.
 From Equation (7),
   $\Delta M_{\rm cold} =  1.52\times 10^{24}$ g
 for the stored mass.
  Given a mass transfer rate typical of systems
  above the period gap $\simeq 2\times 10^{-9}\msunyr$
  we use  Equation (13) to arrive at a recurrence time
  $\sim146$ d.
  The observed recurrence time for outbursts in U Gem is $\sim120$ d,
 and for long outbursts only it is $\sim240$ d 
  (since the outburst pattern basically alternates between long and short 
  outbursts).
  The long outbursts are more relevant as regards
 the recurrence time scale obtained from Equation (13)
   since it is assumed most of the stored mass
    accretes.
     Thus for both WZ Sge and U Gem 
  the theoretical and observed
  recurrence times  for long outbursts
agree to within a factor of $\sim2$,
   indicating both systems
  can have the same scaling  of 
 $\alpha_{\rm cold}$ on binary orbital period.
   Since the orbital periods of the two systems
   are quite different, their values
 of $\alpha_{\rm cold}$ are quite different.

The $e-$folding decay times $t_d \simeq 10$ d  
     for the mass of the accretion disk
  during SO in  V1504 Cyg and V344 Lyr represent $\sim10^3$
orbital time scales in the outer disk $t_{\phi} = 2\pi \Omega^{-1} = $
540 s $m_1^{-1/2} r_{10}^{3/2}$.
  In other words $t_d/t_{\phi} \simeq 10^3$.
  The latest  MRI simulations of accretion disks,
which should be applicable to the outburst state of DNe,
are not yet sophisticated enough to be directly compared
to SOs.
 For instance,  Sorathia et al. (2011)
  perform global magnetohydrodynamic simulations of 
  accretion disks which span a dynamic range of 
  only a factor 
of four in disk radii. They also neglect the vertical dependence
of the gravitational potential, which does not allow the effects
of magnetic buoyancy in reducing the magnetic pressure within the body 
of the disk.  In their simulations they find
    $t_d/t_{\phi} \simeq 10^2$ 
  for the depletion of accretion disk mass.
  Nevertheless their results are quite interesting in 
the context of our work because they show that the Shakura-Sunyaev
$\alpha$ values determined from the local $r$ and $\phi$ components
  of the magnetic field
 are about an order of magnitude lower than those determined 
from the averaged viscous time scale in the outer disk
 (equivalent to the global decay rate of the disk mass), 
  which is the more relevant
    determination in the context of comparison with DN decay rates.
   King, Pringle, \& Livio (2007) noted a factor $\sim10$ discrepancy
between observationally determined $\alpha_{\rm hot}$ values 
    and those calculated numerically from MRI 
  simulations (the MRI values being too small). 
   Although the local shearing box simulations 
  cited by King et al. have
been strongly criticized  because of convergence issues
   (Fromang \& Papaloizou 2007),
the global calculations of
  Sorathia et al. also give
$\alpha_{\rm hot}\simeq 0.01$.
   If the general trend between two methods
of  $\alpha_{\rm hot}$
    determinations
 by Sorathia et al.  are confirmed
  in more sophisticated global MRI calculations with
larger $r_{\rm outer}/r_{\rm inner}$
  and realistic vertical stratification, 
   that may resolve the current apparent 
   discrepancy between theory and observation.

\begin{figure}
\centering
\epsscale{1.0}
\includegraphics[scale=0.415]{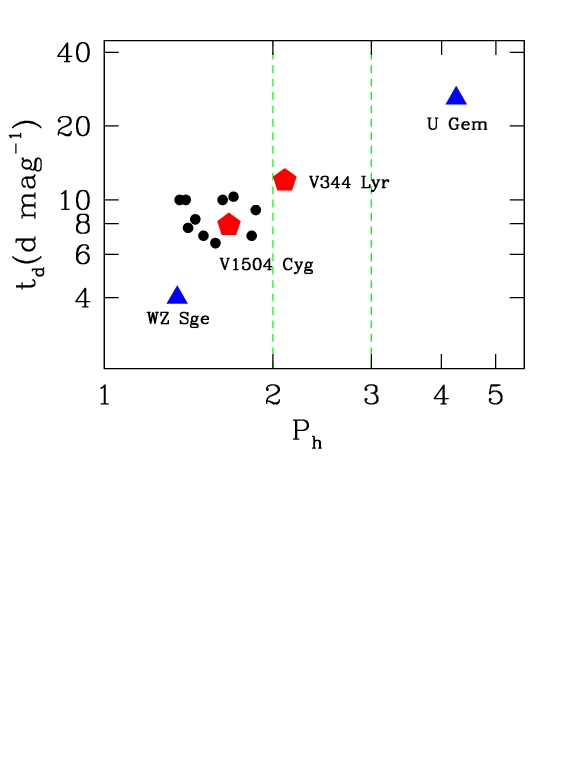}
\vskip -4.75cm
\figcaption{
A comparison of superoutburst slow decay rates
  of other SU UMa stars,
 from Table 2, with
  V1504 Cyg and V344 Lyr.
 Also shown are WZ Sge 
 (initial superoutburst decay rate only)
and U Gem. The vertical dashed
  lines indicate the $2-3$ hr orbital period gap
 for DNe.
\label{fig11}}
\smallskip
\end{figure}

\section{Conclusion}

We have analyzed  long term {\it Kepler} light curves
  of V1504 Cyg, containing 65 outbursts, and V344 Lyr, containing
  64 outbursts.  Each system showed six superoutbursts.
     The findings are:

\smallskip

(1) The NO decays are faster-than-exponential, 
    with most of the deviation occurring near maximum light.
    Near quiescence the decays are close to exponential,
     $\sim$0.7 d mag$^{-1}$ for V1504 Cyg  
    and 
     $\sim$0.6 d mag$^{-1}$ for V344 Lyr.
 These rates are in line with that expected from the Bailey relation. 
  Our values are based on  
       frequency histogram distributions
of only $\sim$50 NOs and should improve as more outbursts are 
accumulated. 

\smallskip

(2)   The NO outburst durations $t_b$ 
    increase by a factor $\sim$1.2$-$1.9
    over the time interval
    between consecutive superoutbursts,
  resetting to a small value after a  superoutburst.
  This is consistent with a monotonic increase of the 
  accretion disk mass with each successive NO.

\smallskip

(3)    The
  quiescent intervals $t_q$ between  normal outbursts 
  show a general trend in both systems of
increasing to a local maximum about 
     mid-way between superoutbursts.
  (There was also an anomalously 
  long quiescence interval at the 
start of   the V344 Lyr light curve,
   potentially the result of a tilted disk.)
    This behavior for $t_q$  
     is inconsistent with
    Osaki's thermal tidal instability 
    model for SOs,
    in which one expects a monotonic increase in $t_q$
   values accompanying the 
  monotonic build-up in disk mass and angular momentum
   leading  to a SO.
If $t_q$ is correlated with the NO triggering radius $r_{\rm trig}$,
then $r_{\rm trig}$ moves outward
 with each successive NO
   only through 
roughly the first half of a
supercycle, and then recedes.
 
\smallskip

(4) The NOs in both systems are asymmetric,
   i.e., $t_{\rm rise}/t_b < 0.5$,
   consistent with ``Type A'' outbursts (Smak 1984).

\smallskip

(5)
  The inference of 
 the
   steep relation  $\alpha_{\rm cold} \simpropto P_h^{4}$
   gives strength to the notion of
   tidal torquing as the 
dominant angular momentum transport mechanism operating
   in quiescent accretion disks in interacting binaries.
 
\smallskip

If one 
considers the slow decay rate of the long, viscous
  outbursts spanning DNe from WZ Sge  to U Gem,
   one infers an overall variation $\simpropto P_h^{1.6}$.
It is unfortunate that only one long outburst in one DN longward
of the $2-3$ hr period gap exhibited enough dynamic variation
in flux during its plateau phase
that a decay rate can be reliably extracted for the slow decay portion.
  It is fortuitous,
  given the general scatter in DN outburst properties
at a given orbital period,
  that
the decay rate between V1504 Cyg and V344 Lyr
   shows a similar law 
   $\simpropto P_h^{1.8}$.
  These dependencies
   are 
 much steeper than the $P_h^{0.25}$
   expected from the previous theoretical expression,
  due to 
 Warner (1995a,
         1995b).
  By starting with the definition for the viscous time scale
in the outer disk during superoutburst, and taking into account 
the scaling for material accumulated during quiescence, we
   derive a more physically motivated formula, 
  and find that
a steep dependence of the quiescent state value of $\alpha$ is
required, $\alpha_{\rm cold}\simpropto P_h^{4.2}$.  In simple terms,
  the dramatic increase in stored quiescent mass with decreasing
orbital period brings about a decrease in the viscous time scale
of matter in the superoutburst disk, which is formed out of redistributed
  gas from the quiescent state.
   Recent work by Patterson
     (2011) 
   on the recurrence time scale in SU UMa
stars for SOs  
   also supports a steep dependence of $\alpha_{\rm cold}$ on $P_h$.
 The long outbursts in systems above the period gap and
superoutbursts in systems below the gap both represent viscous
decays, with the absence of transition waves
in the disk.
   It is somewhat counterintuitive that the recurrence time for 
  these long, viscous outbursts
   varies as $P_h^{-2.6}$, 
  and yet their decay time 
   varies as $P_h^{1.8}$.
  The steepness of the  $\alpha_{\rm cold}(P_h)$  relation
  explains why  accretion disk modelers have 
   had to adopt such small $\alpha_{\rm cold}$ values for the SU UMa stars,
  compared to DNe above the period gap, and also probably  
  accounts for the fact that the extreme mass ratio black hole
   transient systems undergoing limit cycle accretion disk
     outbursts,  like A0620-00,  have such long recurrence times
  for outbursts.

   The {\it Kepler} data have already  provided important constraints
   on the physics of accretion disks, in particular the Shakura-Sunyaev  
  $\alpha$ parameter: C10 showed that an interpolation between $\alpha_{\rm cold}$
  and $\alpha_{\rm hot}$ based on the degree of partial ionization of 
  gas was needed to account for the shoulders in the SOs of V344 Lyr, 
   and in this work we have shown that the scaling of SO decay rate
  with orbital period mandates a steep  dependence of $\alpha_{\rm cold}$  on 
orbital period, enforcing the notion of tidal torquing as being
  the dominant viscosity mechanism in quiescent accretion disks.

\smallskip

  We acknowledge the contributions
  of the entire {\it Kepler}  team.

\def\mnras{MNRAS}
\def\apj{ApJ}
\def\apjs{ApJS}
\def\apjl{ApJL}
\def\aj{AJ}
\def\araa{ARA\&A}
\def\aap{A\&A}
\def\aapl{A\&AL}
\def\pasj{PASJ}


\end{document}